\documentclass{scspaperproc}

\usepackage{latexsym}
\usepackage{graphicx}
\usepackage{mathptmx}

\usepackage{amsmath}
\usepackage{amsfonts}
\usepackage{amssymb}
\usepackage{amsbsy}
\usepackage{amsthm}
\usepackage{dsfont}
\usepackage{comment}
\usepackage{algorithm}
\usepackage{algpseudocode}
\usepackage{thmtools, thm-restate}
\usepackage[pdftex,colorlinks=true,urlcolor=blue,citecolor=black,anchorcolor=black,linkcolor=black,bookmarks=false]{hyperref}

\usepackage{cleveref}
\usepackage{hyphenat}
\newtheoremstyle{scsthe}
{8pt}
{8pt}
{\it}
{}
{\bf}
{.}
{.5em}
{}

\theoremstyle{scsthe}
\newtheorem{theorem}{Theorem}

\newtheorem{corollary}[theorem]{Corollary}

\begin{document}
\newcommand{\Var}{\mathrm{Var}}
\newcommand{\Cov}{\mathrm{Cov}}
%
%
\SCSpagesetup{Ahmed, Rahimian, and Roberts}
\def\SCSconferencename{Annual Simulation Conference}

\def\SCSconferenceacro{ANNSIM'24}

\def\SCSpublicationyear{2024}

\def\SCSconferenceeditors{P.J. Giabbanelli, I. David, C. Ruiz-Martin, B. Oakes and R. C\'{a}rdenas}

\def\SCSconferencedates{May 20-23}

\def\SCSconferencevenue{American University, DC, USA}

\title{Optimized Model Selection for Estimating Treatment Effects from Costly Simulations of the US Opioid Epidemic}



\author[\authorrefmark{1}]{Abdulrahman A. Ahmed}
\author[\authorrefmark{1}]{M. Amin Rahimian}
\author[\authorrefmark{2}]{Mark S. Roberts \vspace{-0.5em}}

\affil[\authorrefmark{1}]{Department of Industrial Engineering, University of Pittsburgh, PA, USA}
\affil[ ]{\textit {aba173@pitt.edu}, \textit{rahimian@pitt.edu}}

\affil[\authorrefmark{2}]{Department of Health Policy and Management, University of Pittsburgh, PA, USA}
\affil[ ]{\textit{mroberts@pitt.edu} \vspace{-2em}}

\maketitle

\section*{Abstract}
Agent-based simulation with a synthetic population can help us compare different treatment conditions while keeping everything else constant within the same population (i.e., as digital twins). 
Such population-scale simulations require large computational power (i.e., CPU resources) to get accurate estimates for treatment effects. 
We can use meta models of the simulation results to circumvent the need to simulate every treatment condition. Selecting the best estimating model at a given sample size (number of simulation runs) is a crucial problem. Depending on the sample size, the ability of the method to estimate accurately can change significantly. In this paper, we discuss different methods to explore what model works best at a specific sample size. In addition to the empirical results, we provide a mathematical analysis of the MSE equation and how its components decide which model to select and why a specific method behaves that way in a range of sample sizes. The analysis showed why the direction estimation method is better than model-based methods in larger sample sizes and how the between-group variation and the within-group variation affect the MSE equation.


\textbf{Keywords:} epidemiological models, treatment effects, model selection, and regression model.

\section{Introduction}
\label{sec:intro1}
\vspace{-1em}
Agent-based modeling (ABM) is a useful tool that helps learn epidemic dynamics. By developing a synthetic population and assigning agents to households, workplaces, schools, and public transit, the epidemic model can be more realistic \cite{cajka2010}. For example, authors in \cite{ferguson2006strategies} develop a large-scale simulation to study treatment conditions of an Influenza outbreak across the UK and the US. The simulation was informative for decision-makers to know the efficacy of specific policies (e.g., school closure, vaccine stockpiling, workplace restrictions, etc.). Building on this work, a nationwide simulation is conducted to test Influenza vaccination policies as shown in \cite{chao2010flute}, where the model is calibrated with historical data. Scaling up the simulation size, authors in \cite{parker2011globalscale} develop a simulation that can have billions of agents to simulate global-scale epidemics. In a variation to the previous models, authors in \cite{tuite2010MCMC} build a Markov chain Monte-Carlo simulation model calibrated on historical data of lab-confirmed cases for H1N1 Influenza. Utilizing previous work on ABM simulation paved the way to create a generalized population-scale simulation to study epidemic dynamics called FRED (A Framework for Reconstructing Epidemiological Dynamics) \cite{grefenstette2013fred}. Using FRED, the authors in \cite{lukens2014fluh1n1} create an ABM linked with an equation-based within-host model for Influenza. School closure policies were also studied during the Influenza outbreak \cite{potter2012school}. For a different epidemic, authors in \cite{liu2015measle} study the different vaccination policies for Measles using FRED software, this paper was pivotal in raising awareness about the significance of vaccination and its impact on epidemic outbreaks. Outside of the commonly studied epidemics, the authors in \cite{krauland2020cardio} focus on cardiovascular disease and its mortality. They show the utility of FRED for understanding disease risk and the effects of large-scale interventions \cite{krauland2020cardio}.

\subsection{Related Work}
\vspace{-1em}
Authors in \cite{ahmedbhi} show different methods to estimate the treatment effects of Opioid Use Disorder (OUD) interventions by allocating simulation samples to unknown treatments. 
Our problem is related to the ranking and selection (R\&S) problem, where the goal is to select a subset of models out of a large number of models based on a defined performance \cite{law2007simulation}. Authors in \cite{goldsman1998statistical} provide a review of the R\&S problem in simulation contexts. It is different from simulation optimization, where the goal is to search a parameter space efficiently. R\&S methods aim to evaluate all models from a defined set exhaustively \cite{futheory2002optimization}. Authors in \cite{chen2008efficient} study a variation of R\&S with the optimal allocation of samples. While authors in \cite{boesel2003cleanup} address the problem of R\&S by developing a method to exclude the inferior models from the best-selected subset models. Their work was based on \cite{chen2000ocbda}, where they addressed the R\&S problem from the perspective of allocation of computational budget to more critical models to increase the probability of correct selection under a framework defined as optimal computing budget allocation (OCBA). They use mean and variance in their allocation method, which differs from previous methods that used variance alone \cite{rinott1978variance}. The goal of OCBA is to increase the selection probability of the best method for a specific computational budget. Authors in \cite{peng2016dynamic} propose a Bayesian procedure for OCBA, and  \cite{shi2022dynamic} address the problem with a defined finite-budget rule where under finite-budget simulation, the procedure will increase the sampling ratio from less critical models and decrease the sampling ratio for more critical models. 



\subsection{Main Contribution}
\begin{figure}
    \centering
    \includegraphics[scale=0.7]{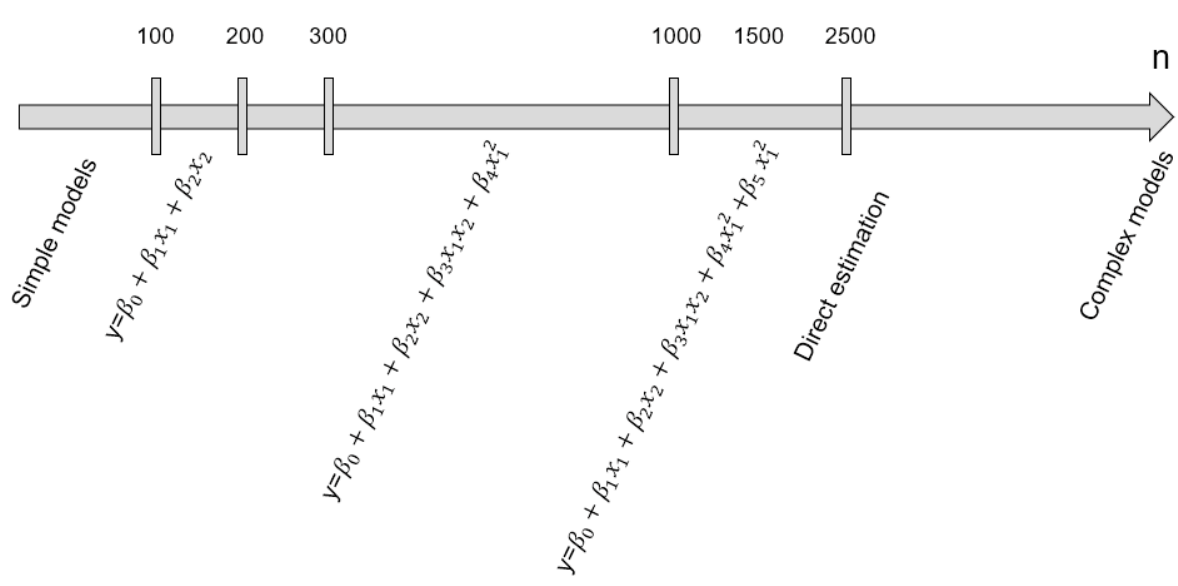}
    \caption{{\bf Model selection in costly sample regimes.} This figure shows which model will have the lowest MSE, given the sample size. The arrow points in the direction of increasing sample size, and at each interval, the equation specified is for the model that achieves the least MSE. Notice the increasing complexity of the optimal model with increasing sample size. With a large enough number of samples directly estimating each treatment condition is optimum.}
    \label{fig:n_axis}
\end{figure}

\vspace{-1em}
Consider that we want an accurate estimate for a specific treatment effect from a large-scale simulation. This can not be conducted with a few simulation runs due to the randomness contained in the simulation. And at the same time this will require an amount of computational computer (i.e., central processing unit resources required to get an accurate estimate for the treatment effect). Therefore, we need to look for a suitable method to estimate the treatment effect. \Cref{fig:n_axis} explains the model selection problem, i.e., at each sample size, what model should we use? The models shown are based on regression equations. However, it starts from the simplest one, i.e., composed of only two covariates with intercept, to contain a quadratic term until including a cubic term in the regression equation. The intuition behind these different models is that the more terms, the more complex the model is to explore the model selection carefully. In addition to this, for comparison, we include the direct-estimation method, which simply calculates the average of the current samples as a benchmark to know when using a model becomes useless.
In addition to the empirical results, we provide a mathematical analysis to understand what are the main components behind model selection. The result shows that while sample size is the main consideration in the choice of simple or complex models, model selection is also affected by properties of the information environment, e.g., within-group variability of the conditions that we are estimating, between-group variation among the mean treatment effects, and the number of levels at which an intervention is applied. For example, model-based methods converge to a non-zero MSE as the sample size goes to infinity due to their bias. However, MSE for the model-free method that directly estimates the treatment effects converges to zero as the sample size goes to infinity. Our theoretical analysis in \Cref{sec:math} sheds light on this issue as $n\to\infty$, the variance term vanishes, and the bias term is dominant, which means that more complex models with less bias are preferred in large sample regimes. 

This paper is structured in five sections. In \Cref{sec:fred}, we introduce the FRED software briefly and its features. In \Cref{sec:methods}, we discuss the different methods we will demonstrate to estimate treatment effects. In \Cref{sec:math}, we provide a mathematical analysis of the behavior of two types of methods in estimating treatment effects (model-based, using linear regression, and model-free, direct estimation). We conclude our paper in \Cref{sec:conclude} with concluding remarks and future directions.

\section{FRED Simulation Software}
\label{sec:fred}
\vspace{-1em}
FRED (Framework for Reconstructing Epidemiological Dynamics) is an agent-based, open-source software that simulates epidemics' temporal and spatial behaviors. The Public Health Dynamics Laboratory (PHDL) at the University of Pittsburgh School of Public Health was behind the development of the FRED software. Initially, FRED was developed to study the epidemic dynamics; however, FRED has shown the potential to give insights into public health and intervention studies. One of the significant features of FRED is that its synthetic population is built on the true US Census \cite{guclu2016agent}.

\paragraph{Synthetic Population:} One of the key features of FRED is its synthetic population, where FRED represents every individual in every specific location explicitly. FRED makes use of the US synthetic population from RTI international \cite{rti}. The synthetic population is assigned to specific geographically allocated places, i.e., each resident is assigned to a specific household, students are assigned to schools, and workers are assigned to workplaces. The specific geographic assignment for agents will also mirror the real spatial distribution of the area and the distance traveled by the agent to their assigned place (e.g., school, workplace, household, etc.). Each agent has its own demographic and socioeconomic information (e.g., race, age, sex, employment, etc.) and specific locations for their business (e.g., school, workplace, household, etc.).

\paragraph{Discrete-time Simulation:} FRED conducts discrete-time simulation with a step size of a day; each day (i.e., simulation step), each agent can meet other agents who share the same geographic location. For example, an agent interacts with other agents within the same household. If the agent is infected with a disease, there is a defined probability that its relatives (i.e., household residents) will get infected by that disease. Each infection transmission event is recorded in the software, which can be used to evaluate control measures. Each agent has the option to change its daily activity, e.g., not to go to the workplace on a specific day or travel from the current location.

\paragraph{Agent Model:} Each agent has its own demographic features (e.g., age, race, sex, employment, etc.), location for activities (e.g., school, workplace, household, neighborhood, etc.), and health-related information (e.g., staying at home when sick, probability of getting a vaccine). In addition, FRED allows us to keep the demographic features constant or not. For example, if the demographic features change is enabled, then the agent's age will change and could affect their employment and other aspects. Adult agents that reach the working age are assigned to workplaces; similarly, children that reach school age are assigned to schools. Infants are assigned to the same household as their parents, and if an agent dies, it is removed from the synthetic population. Agents have options in their health-related decisions (e.g., agents can make decisions like taking a vaccine or not and staying home when sick or not).

\paragraph{Disease Model:} FRED supports the spreading of one or more infectious diseases. Each disease development is ruled by precise parameters for contact, transmission, and natural history. From an agent's perspective, the agent is expected to follow a model-specified path. For example, The agent will pass through the classic Susceptible, Infected, and Recovered (S-I-R) stages where the agent will move susceptible to infection based on the transmission rate and contact rate (e.g., if the agent is within a school that has disease-spreading agent will have a higher transmission rate compared to non-school agents). FRED specifies the contact details also where the transmission rate between students will be higher than teachers, even if the teachers were at the same school. FRED considers every contact an independent transmission opportunity (e.g., if an infected student meets the same susceptible student multiple times, each time is considered an independent transmission opportunity). 
Moreover, FRED supports the spreading of multiple strains in the same population where the intensity and trajectory of every strain is defined by the model developer.

\section{Model Selection in Costly Sample Regimes}
\label{sec:methods}
\vspace{-1em}

Due to the randomness contained in the simulation, we can not get an accurate estimate of treatment effects easily (i.e., small confidence interval width). Therefore, we are required to run the simulation multiple times until we get a specified accuracy for treatment effects. In this section, we will discuss possible methods we can use to get an accurate estimate of treatment effects.

Assume that we have $L$ treatment conditions (e.g., from applying an intervention at $L$ different levels), and samples in each condition are independent and normally distributed with mean $y_{\ell}$ and variance $\sigma^2$, $\ell =  1, \ldots, L$. We denote the $i$-th sample in the $\ell$-th condition by $y_{\ell i}$, which is normally distributed random sample, $y_{\ell i} \sim {N}(y_{\ell},\sigma^2)$. Our goal is to construct estimates of the means in each treatment condition $\hat{y}_{\ell}$ to minimize the following expected squared loss:
\begin{align}
    MSE = \sum_{\ell=1}^{L}(\hat{y}_{\ell} - y_{\ell})^2.
\end{align}
\paragraph{Direct Estimation: } The simplest solution to this solution uses sample means
$$\hat{y}_{\ell} = \frac {\sum_{i=1}^{n/L} y_{\ell i}}{n/L},$$ and achieves $MSE = L^2 \sigma^2 /n$. Here we have assumed that $n$ simulation samples are allocated equally across the $L$ conditions. Samples means are unbiased and simple to estimate but the resultant MSE may not be optimized for costly simulation regimes where $n$ is small.

\paragraph{Using models to learn across treatment conditions: } In a costly simulation regime where sample size $n$ is low, we can model the effect of treatments such that a new batch of samples for treatment effect A also allows us to improve the accuracy of estimates for treatment effect B. The idea is that instead of focusing on each treatment effect case by case, we can look into the adjacent treatment effects as a whole sample space. With the help of the regression equation, the current estimate of a specific treatment effect will be updated not only from its new sample batch but also from neighboring sample batches. This could help reduce the required number of samples to achieve a pre-specified accuracy (e.g., achieving a specific confidence interval (CI) width).

\paragraph{Bias-Variance trade-off: } Using models allows us to reduce the variability of our estimates by making better use of the available samples across all conditions, but it comes at the cost of the increased bias of model-based estimation. The bias-variance trade-off is one of the oldest known statistical problems describing the trade-off between the complexity of the model and its accuracy in prediction. Consider if we have $y=f(x)+\varepsilon$ where $E(\varepsilon)=0$ and $Var(\varepsilon)=\sigma^2_\varepsilon$, define a regression fit $\hat{f}(x)$ for input $X=x$ the squared loss can be defined as:
\begin{align}
    Error(x) &= E[(y-\hat{f}(x))^2|X=x] \nonumber
    \\
    &= (E[\hat{f}(x)]-f(x))^2+E[\hat{f}(x)-E[\hat{f}(x)]]^2+ \sigma^2_\varepsilon \nonumber
    \\
    &= Bias^2 + Variance + noise
\end{align}
Where the first term is the bias, which measures how much the average of the estimate is different from the true mean, the second term is the variance of the estimate, and the third term is the noise term \cite{hastie2013introduction}. In Section \ref{sec:math}, we explain this trade-off for estimating the treatment effects of an intervention that can be applied at $L$ levels. Our results clarify the choice between directly estimating the $L$ treatment conditions using sample means or using a linear regression where the $L$ conditions are modeled as $L$ level of a factor $x$ that can take values $x_{\ell} = \ell$, $\ell = 1, \ldots, L$. The $x_{\ell} = \ell$ encoding of the levels is arbitrary and can be optimized to improve MSE if one has some prior knowledge of the population means $y_{\ell}$ at each level. 

\section{Empirical Results}
\label{sec:result}
\vspace{-1em}
\paragraph{OUD model:} The Opioid Use Disorder (OUD) model is developed to understand the OUD epidemic nationwide. The rise of drug overdose and opioid use disorder is a public health concern in the US currently. The current OUD wave is part of a decades-long trend stressing the importance of studying OUD dynamics \cite{jalal2018opioiddynamics}. The PHDL developed the OUD model we use in this paper at the University of Pittsburgh based on data provided by the Centers for Disease Control and Prevention (CDC) within a sponsored project by the CDC. The OUD model is updated monthly, where the OUD deaths are reported at specific locations (as a more informative way for the researchers). The results used in this paper were conducted for Allegheny County, PA.

\begin{figure}[h]
    \centering
    \includegraphics[scale=0.39]{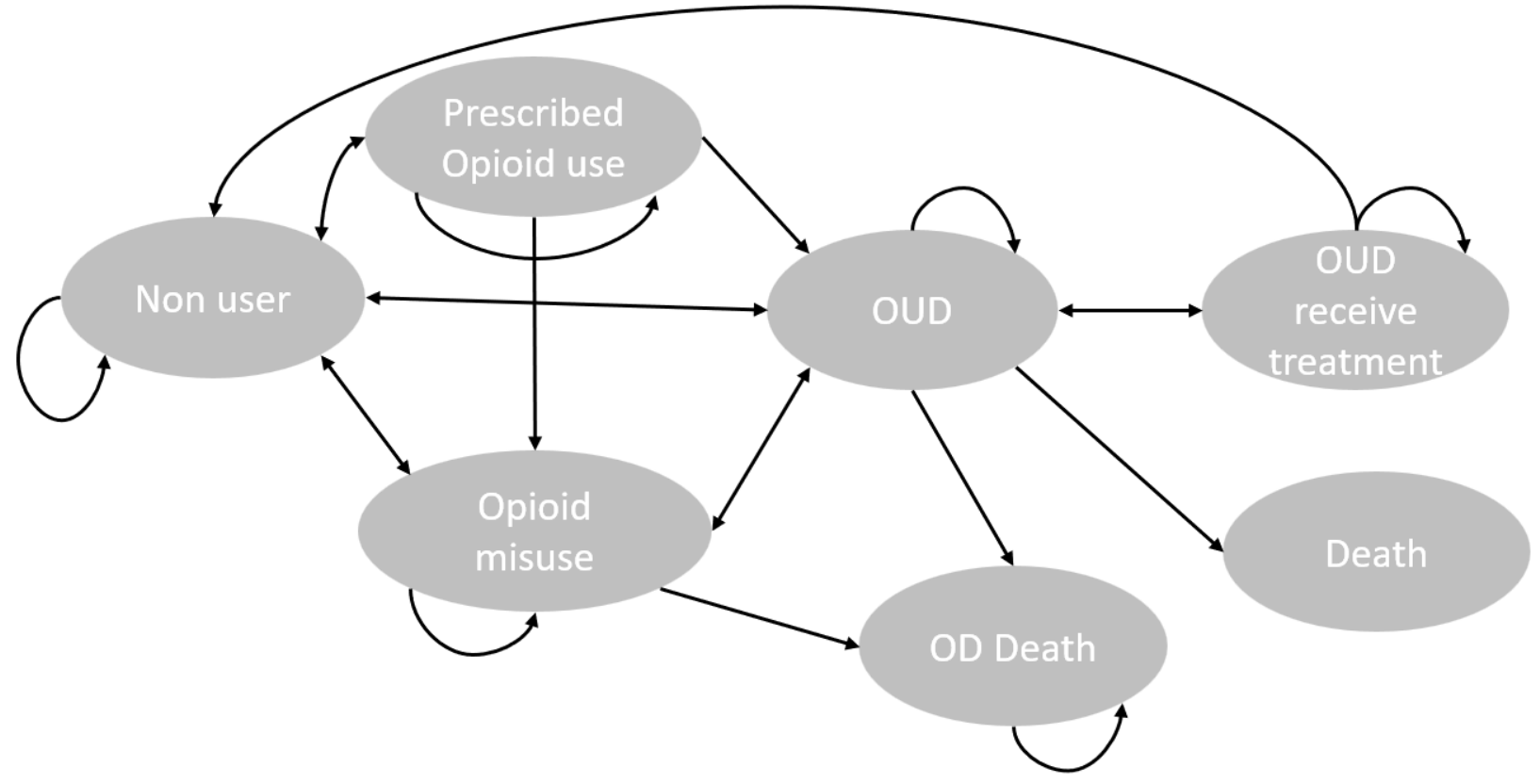}
    \caption{State transition diagram for the OUD model.}
    \label{fig:oudFlow}
\end{figure}

\paragraph{Two interventions: }Consider a problem where we have two interventions, Buprenorphine and Naloxone, which are used to mitigate OUD harms. Buprenorphine is a medication provided for OUD patients within treatment to move from a misuser state to a non-user. Naloxone is a medication used to reverse the effect of an opioid overdose (i.e., an overdose antidote). Each factor has five levels, representing the amount of the factor (medication) available in a specific location. 
This will result in 25 treatment conditions (Combining the two factors levels). We selected these two medications specifically as they are effective in the treatment process from the OUD and reduce the number of deaths from OUD overdose.
Authors in \cite{ahmedbhi} show results for model-free and model-based methods for the two-factor problem. 

\subsection{Comparing MSEs by varying model complexity over sample size}
\vspace{-1em}

\begin{figure}[htb]
{
\centering
\includegraphics[scale=0.5]{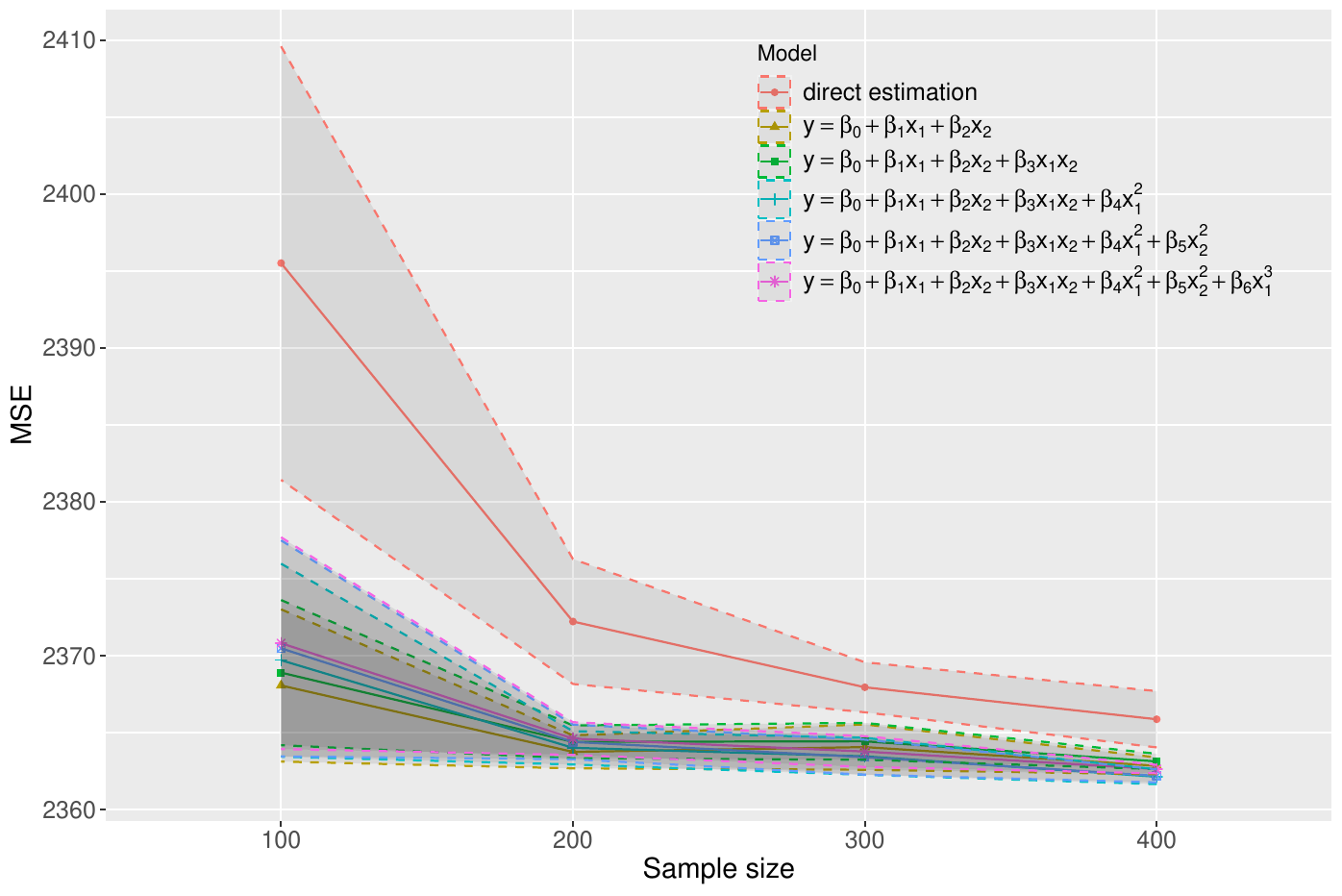}
\caption{The MSE values for different models compared to sample sizes of 100, 200, 300, and 400, respectively, where it show that simpler model-based methods are better than higher terms model-based methods and model-free methods.}
\label{fig:smallsample}
}
\end{figure}

\begin{figure}[htb]
{
\centering
\includegraphics[scale=0.5]{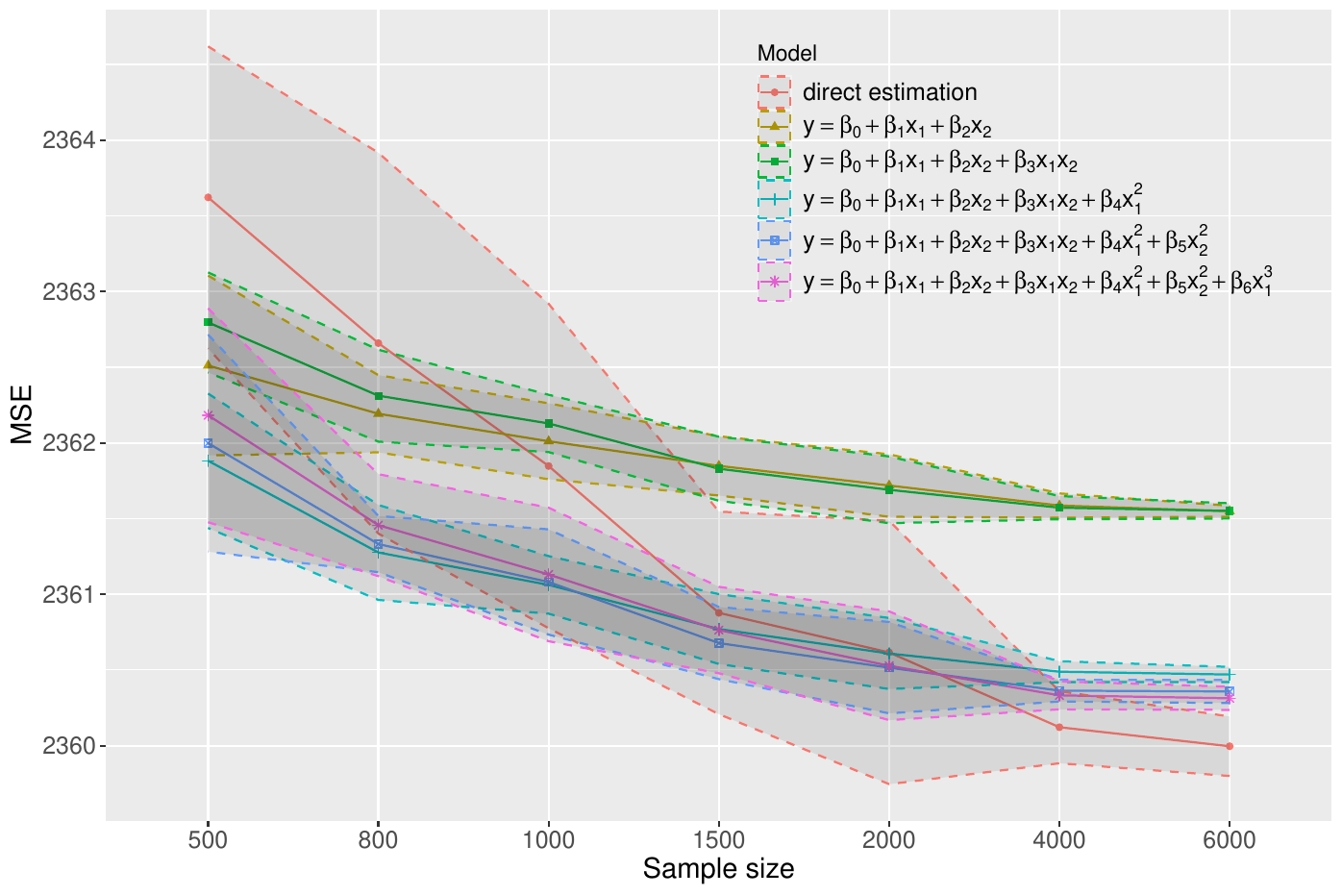}
\caption{The MSE values for different models compared to sample sizes from 500 to 6000 where show that model-based methods perform well until a specific sample size that the model-free method supersedes in performance.}
\label{fig:largesample}
}
\end{figure}

We have seen that model-based methods performed better than model-free methods in terms of required simulation runs for pre-specified CI width, as shown in \cite{ahmedbhi}. Consider another aspect: how will each method perform given a specific sample size? To set up this problem, we selected a range of sample sizes starting from 100 to 6000 simulation runs. At each sample size, we calculate the MSE value for each method to evaluate their performance. In addition to the two regression models demonstrated \cite{ahmedbhi}, we explored extra regression models to evaluate better the performance of model-based methods over different sample sizes.
Model1, model2, model3, model4, and model5 represented by equations \ref{eq:reg2}, \ref{eq:reg3}, \ref{eq:reg4}, \ref{eq:reg5}, and \ref{eq:reg6} respectively. All these methods are defined under the category of model-based methods. For comparison, we will add a direct-estimation method as an example of the model-free method performance. \Cref{fig:smallsample} shows the result for small sample sizes, and \Cref{fig:largesample} shows the result for large sample sizes. As we can see, at the beginning, the model-free method was way too high than any model-based method in MSE value; however, as the sample size grows, the model-free starts to get a lower MSE value until at 4000 sample size, where the model-free beats the model-based methods. In the following section, we will show a mathematical explanation for the performance of the two types of methods.

\begin{align}
    y &= \beta_0+\beta_1 x_1
    \label{eq:reg2} \\
    y &= \beta_0 + \beta_1 x_1 + \beta_2 x_2 + \beta_3 x_1x_2 \label{eq:reg3}
    \\
    y &= \beta_0 + \beta_1 x_1 + \beta_2 x_2 + \beta_3 x_1x_2 + \beta_4 x_1^2 \label{eq:reg4}
    \\
     y &= \beta_0 + \beta_1 x_1 + \beta_2 x_2 + \beta_3 x_1x_2 + \beta_4 x_1^2 + \beta_5 x_2^2 \label{eq:reg5}
     \\
      y &= \beta_0 + \beta_1 x_1 + \beta_2 x_2 + \beta_3 x_1x_2 + \beta_4 x_1^2 + \beta_5 x_2^2+ \beta_6 x_1^3 \label{eq:reg6}
\end{align}

\section{Theoretical Analysis}
\label{sec:math}
\vspace{-1em}
As we have seen in the previous section, in some cases, the model-based methods perform better than the model-free ones, but as the sample size becomes very large, the model-free method becomes better in terms of MSE. What was the reason behind that change with respect to the sample size? To evaluate the performance of the proposed methods, we analyze and compare the MSEs for direction estimation and a linear regression model with one factor.
Consider a sample size $n$ and $L$ treatment conditions with mean effects $y_{\ell}$ which we aim to estimate using samples $y_{{\ell}i}$, $\ell = 1,2,\ldots, L$ which are independently and normally distributed with mean $y_{\ell}$ and variance $\sigma^2$, i.e., within-group variation. We encode the $L$ treatment groups as $L$ levels of a factor, $x_{\ell}, \ell = 1, 2, \ldots, L$. We use the arbitrary encoding $x_{\ell} = \ell$ and further associate $x_{\ell i}= \ell$, $\ell = 1,2,..,L$ with the $i$-th observation in the $\ell$-th group. Moreover $\bar{x}= \sum_{\ell = 1}^{L}x_{\ell} =  (L+1)/2$. We define the difference between level means $y_l$ as a between-group variation.

\vspace{0.5cm}
\begin{restatable}{thm}{mse}
\label{thm:mse}
Consider a sample size of $n$ with $L$ levels and assume that samples are allocated equally across the levels ($n/L$ samples to each level). A model-based estimate of the $L$ treatment effects using a linear function with least-squares fit to the observed samples $\hat{y_l}=\hat{\alpha}+\hat{\beta}x_l$ and $x_{\ell} = \ell, \ell = 1,2,\ldots,L$ of the levels, gives the following MSE:\\
\begin{equation}
\mbox{MSE} = \frac{L\sigma^2}{n} \left( \frac{  2(1/3L^2+1/2L+1/6)}{(1/12)(L^2-1)}
    +1 \right)
    +\frac{(2L^2+3L+1)}{6L}  \left( \frac{\sum^L_{\ell=1} (\ell-\frac{(L+1)}{2})(y_{\ell}-\bar{y})}{(1/12)(L^2-1)} \right)^2  \nonumber
\end{equation}
Where the MSE of the model-based method depends on the within-group variation and between-group variation.

\end{restatable}

Theorem \ref{thm:mse} shows that the model-based method MSE is comprised of three components: the within-group variation, the between-group variation, and the number of levels. As within-group variation $\sigma^2$ increases, the ability of the model-based method to estimate the true level mean $y_l$ becomes worse. Similarly, as between-group variation $\rho^2 = \left( \frac{\sum^L_{\ell=1} (\ell-\frac{(L+1)}{2})(y_{\ell}-\bar{y})}{(1/12)(L^2-1)} \right)^2$ increases, this would affect the performance of the model-based method badly. In addition, the number of levels $L$ affects the value of MSE equation for the model-based method as it appears on each term of the equation.
In contrast, for the case of the model-free method, its MSE depends only on the within-group variation $\sigma^2$ and the number of levels $L$. \\ 
This could explain why the model-free method, after a specific sample size (in \Cref{fig:largesample}), its MSE becomes better than model-based methods. This shows that the performance of the model-based method is tied to within-group variation and between-group variation parts mainly.


Consider the critical sample size at which MSE for direct estimation, $MSE = L^2 \sigma^2 /n$ equals the MSE for regression-based estimates derived in \Cref{thm:mse}: 
\begin{equation}
n^{\star} = \sigma^2 \left( L - 1 - \frac{2(1/3L^2+1/2L+1/6)}{(1/12)(L^2-1)} \right)
\frac{6L^2}{2L^2+3L+1}  \rho^{-2}
\label{eq:nstar}
\end{equation}
 This sample size establishes the critical region above which direct estimation outperforms the regression-based estimates; therefore, when the sample size is smaller than $n^*$, then it is beneficial to use the model-based method, while if the sample size is greater than $n^*$, then it is beneficial to directly estimate the treatment effects. Similar to \Cref{thm:mse}, $n^{\star}$ is dependent on the same factors of the information: it is increasing in within-group variation $\sigma^2$ and the number of levels ($L$), and decreases with increasing between-group variation $\rho^2$.

\begin{figure}
    \centering
    \includegraphics[scale=0.6]{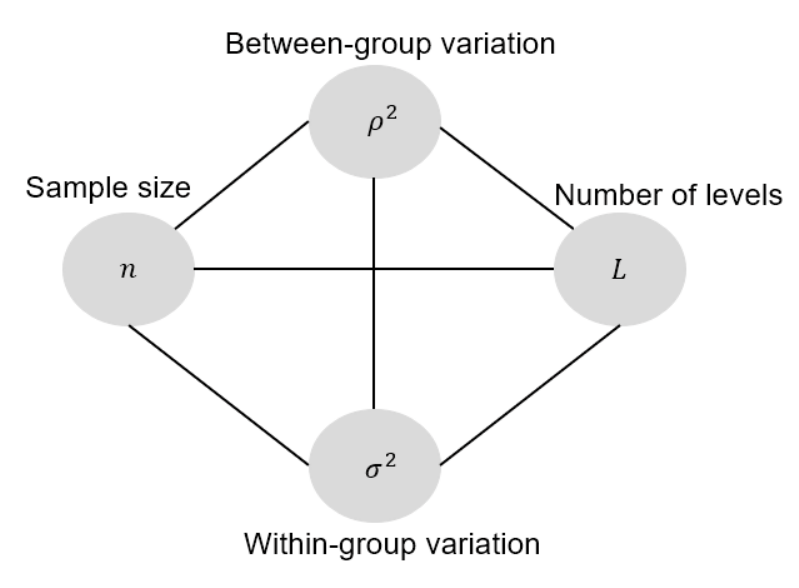}
    \caption{\small {\bf The main components that decide the MSE equation and affect the choice of a model.} This figure shows the main variables that comprise the MSE value. The first one, $\rho^2$, is the between-group variation where its increase will affect the bias part of the MSE equation for the model-based method. The second variable (in clockwise order) $L$ is the number of levels, though, in this problem, we defined a fixed number of levels, but the idea generalized as $L\to\infty$ and \Cref{eq:nstar} for $n^{\star}$ is increasing in $L$ for large $L$. In those regimes, we prefer to use model-based methods over a broader range because the within-group variability term dominates the MSE equation as $L\to\infty$. The third factor affecting our model selection is $\sigma^2$ or within-group variability; with increasing $\sigma^2$, we prefer to use more and more complex models to minimize MSE. Lastly, $n$ is the sample size which is the most critical factor in optimizing model selection (\Cref{fig:n_axis}). As $n \to \infty$, the variance terms go to zero, and the bias part in the model-based MSE equation becomes dominant, at which point direct estimation is preferred (there are no advantages in the use of models in large sample regimes).}
    \label{fig:tetrahedron_var}
\end{figure}
\section{Conclusion}
\label{sec:conclude}
\vspace{-1em}
Estimating treatment effects in a large-scale simulation is a computationally exhaustive task. The straightforward method of brute force can be applied in small-size simulations but does not apply to larger simulation models. Therefore, we explored model-based methods showing different regression models and their performance compared to the model-free method. We demonstrated the methods' performance given different sample sizes to provide more analysis for our approach. 
Furthermore, we provided a mathematical analysis to explain why model-free is better than model-based in larger sample sizes. The analysis shows that the MSE equation for the model-based method depends on the between-group variation and within-group variation, which explains why model-based methods perform better at specific sample sizes than model-free methods in terms of MSE and vice versa. \\
This work can be extended by changing the labeling method; in this paper, we defined the levels as $1,2,3, ..., L$, and we got $\rho^2$ as the weighted sum for between-group variation. However, defining the levels more wisely will get us an unweighted sum of between-group variation. Moreover, the extension can be done by exploring estimation methods that can reach better bias-variance trade-off, in addition, incorporating spatial data could help in understanding better the OUD model dynamics (e.g., the Gaussian process showed potential in learning epidemic dynamics with spatial data \cite{ahmed2023inferring}). Specifically, integrating spatial data could help the model in learning the socio-economic details, which, in turn, gives a more accurate estimate of the treatment effect. We intend to explore this path as a future research direction.


\section*{Data and Code}
For reproducibility, we supported our study with code (https://github.com/abdulrahmanfci/model-selection). However, we are not able to share detailed data about the OUD model for contractual reasons.

\section*{ACKNOWLEDGMENTS}
This research was funded by contract 75D30121C12574 from the Centers for Disease Control and Prevention. The findings and conclusions in this work are those of the authors and do not necessarily represent the official position of the Centers for Disease Control and Prevention.

This research was partly supported by the University of Pittsburgh Center for Research Computing, RRID:SCR\textunderscore022735, through the resources provided. Specifically, this work used the HTC and VIZ clusters, which are supported by NIH award number S10OD028483.

\appendix

\section{Proof of Theorem 1} 
\subsection{MSE in the model-based case}
Consider a problem with a sample size of $n$ and a number of levels $L$ where levels are $1,2,3,..., L$. Each level has an equal amount of samples (i.e., $n/L$), and all levels have the same variance (homoscedasticity). A model-based method is given as $\hat{y_{\ell}}=\hat{\alpha}+\hat{\beta}x_{\ell}$ to estimate the true $y_{\ell}$ where $y_{\ell}$ is the average of $y_{\ell i}$ at level $\ell$, also $\bar{\Bar{y}}=\frac{1}{n}\sum^L_{\ell=1}\sum^{n/L}_{i=1}y_{\ell i}$, $\Bar{y}=\frac{1}{L}\sum^L_{\ell=1}y_{\ell}$ and $\hat{\alpha}=\bar{\Bar{y}}-\hat{\beta}\Bar{x}$ and $\hat{\beta}=\frac{\sum^L_{\ell=1} \sum^{n/L}_{i=1} (x_{\ell i}-\Bar{x})(y_{\ell i}-\bar{y})}{\sum^L_{\ell=1} \sum^n_{i=1} (x_{\ell i}-\Bar{x})^2}$. The $\hat{\alpha}$ and $\hat{\beta}$ are defined using the least-squares fitting equation. \\
The mean squared error (MSE) for the model-free method is defined as:
\vspace{-0.5em}
\begin{equation}
    MSE = E[\sum^L_{\ell=1}(\hat{y_{\ell}}-y_{\ell})^2],
    \vspace{-0.5em}
\end{equation}  
where $\hat{y_{\ell}}$ is the estimator for the true $y_{\ell}$ and $y_{\ell}$ is the true mean for $y_{\ell i}$ values (i.e., death estimates at level $l$). The MSE term can be expanded as follows:

\begin{align} \label{eq:MSE}
     MSE = E[\sum^L_{\ell=1}(\hat{y_{\ell}}-y_{\ell})^2] \nonumber
     &= E[\sum^L_{l=1}(\hat{y_{\ell}}-E[\hat{y_{\ell}}]+E[\hat{y_{\ell}}]-y_{\ell})^2] \nonumber
     \\
     &= E[\sum^L_{\ell=1}( (\hat{y_{\ell}}-E[\hat{y_{\ell}}])^2+ 2(\hat{y_{\ell}}-E[\hat{y_{\ell}}])(E[\hat{y_{\ell}}]-y_{\ell})+(E[\hat{y_{\ell}}]-y_{\ell})^2)] \nonumber
     \\
     &= \underbrace{\sum^L_{\ell=1}(E[\hat{y_{\ell}}^2]-E[\hat{y_{\ell}}]^2)}_{\mbox{variance}}+\underbrace{\sum^L_{l=1} (E[\hat{y_{\ell}}]-y_{\ell})^2}_{\mbox{bias}} \;\;\;\;\;\;\;\;\; \mbox{(by linearity of expectations)} \nonumber
     \\
     \intertext{Working on the variance term} 
      \sum^L_{\ell=1}(E[\hat{y_{\ell}}^2]-E[\hat{y_{\ell}}]^2)
      &= \sum^L_{\ell=1}( \Var(\hat{\alpha})+x_{\ell}^2(\Var(\hat{\beta}))) \nonumber
      \\
      &= \sum^L_{\ell=1} \underbrace{\Var(\bar{\Bar{y}})}_{
      = {\sigma^2}/{n}}-2\Bar{x}\; \Cov(\bar{\Bar{y}},\hat{\beta})+(x_{\ell}^2+\Bar{x}^2) \Var \left( \frac{\sum^L_{\ell=1} \sum^{n/L}_{i=1} (x_{\ell i}-\Bar{x})(y_{\ell i}-\bar{y})}{\sum^L_{\ell=1} \sum^{n/L}_{i=1} (x_{\ell i}-\Bar{x})^2} \right) \nonumber
      \\
      \intertext{Working on the second term}
      &\Cov(\bar{\Bar{y}},\hat{\beta}) = \Cov \left(\frac{1}{L}\frac{L}{n} \sum^L_{\ell=1}\sum^{n/L}_{i=1} y_{\ell i}, \; \frac{\sum^L_{\ell=1} \sum^{n/L}_{j=1} (x_{\ell j}-\Bar{x})(y_{\ell j}-\bar{y})}{\sum^L_{\ell=1} \sum^{n/L}_{i=1} (x_{\ell i}-\Bar{x})^2} \right) \nonumber
      \\
      &= \frac{\displaystyle\sum^L_{\ell=1} \sum^{n/L}_{i=1} \sum^{n/L}_{j=1} (x_{\ell j}-\Bar{x})\Cov \left(y_{\ell i}, y_{\ell j}  \right)}{\displaystyle n \sum^L_{\ell=1} \sum^{n/L}_{i=1} (x_{\ell i}-\Bar{x})^2} = 0 \nonumber \;\;\;\;\;\;\; \mbox{( $\displaystyle\sum^L_{\ell=1} \sum^{n/L}_{j=1} (x_{\ell j}-\Bar{x}) = 0$ \& $\Cov (y_{\ell i}, y_{\ell j}) = \sigma^2 \mathds{1}_{i=j}$)}
      \\
      \intertext{Working on the third term}
      & (x_{\ell}^2+\Bar{x}^2) \Var \left( \frac{\sum^L_{\ell=1} \sum^{n/L}_{i=1} (x_{\ell i}-\Bar{x})(y_{\ell i}-\bar{y})}{\sum^L_{\ell=1} \sum^{n/L}_{i=1} (x_{\ell i}-\Bar{x})^2} \right) = 
      (x_{\ell}^2+\Bar{x}^2)\frac{\sum^L_{\ell=1} \sum^{n/L}_{i=1} (x_{\ell i}-\Bar{x})^2 \Var(y_{\ell i})}{[\sum^L_{\ell=1} \sum^{n/L}_{i=1} (x_{\ell i}-\Bar{x})^2]^2} \nonumber
      \\
      &= (x_{\ell}^2+\Bar{x}^2)\frac{\sigma^2}{\sum^L_{\ell=1} \sum^{n/L}_{i=1} (x_{\ell i}-\Bar{x})^2} \nonumber
      \\
      \sum^L_{\ell=1}(E[\hat{y_{\ell}}^2]-E[\hat{y_{\ell}}]^2) &=\sum^L_{\ell=1} \frac{\sigma^2}{n}+(x_{\ell}^2+\Bar{x}^2)\frac{\sigma^2}{\sum^L_{\ell=1} \sum^{n/L}_{i=1} (x_{\ell i}-\Bar{x})^2} 
      \;\;\;\;\;\;\;\;\;\;\;\;\;\;\;\;\;\;\;\;\;
      \mbox{(Putting the two terms together)}
      \nonumber
      \\
      &= \sum^L_{\ell=1} \frac{\sigma^2}{n \: \sum^L_{\ell=1} \sum^{n/L}_{i=1} (x_{\ell i}-\Bar{x})^2} \left( \sum^L_{\ell=1} \sum^{n/L}_{i=1} (x_{\ell i}-\Bar{x})^2+ n(x_{\ell}^2+\Bar{x}^2) \right) \nonumber
      \\
      &= \sum^L_{\ell=1} \frac{\sigma^2 (\sum^L_{\ell=1} \sum^{n/L}_{i=1}x^2_{\ell i}+nx^2_{\ell})}{n \: \sum^L_{\ell=1} \sum^{n/L}_{i=1} (x_{\ell i}-\Bar{x})^2} \nonumber 
      \\
      \intertext{To simplify the variance term}
      &= \frac{\sigma^2}{n} \sum^L_{\ell=1} \frac{ (\sum^L_{\ell=1} \sum^{n/L}_{i=1}x^2_{\ell i}+nx^2_{\ell})}{\sum^L_{\ell=1} \sum^{n/L}_{i=1} (x_{\ell i}-\Bar{x})^2}
      = \frac{\sigma^2}{n} \sum^L_{\ell=1} \frac{ n/L(L(L+1)(2L+1)/6)+nx^2_{\ell}}{\sum^L_{\ell=1} \sum^{n/L}_{i=1}x^2_{\ell i}-n\Bar{x}^2}
      \nonumber \\
      &= \frac{\sigma^2}{n} \sum^L_{\ell=1} \frac{ n/L(L(L+1)(2L+1)/6)+nx^2_{\ell}}{n/L(L(L+1)(2L+1)/6)-n\frac{L^2}{4}}
      = \frac{\sigma^2}{n} \frac{ \sum^L_{\ell=1} (1/3L^2+1/2L+1/6+x^2_{\ell})}{7/12L^2+1/2L+1/6}
      \nonumber
      \\
      &= \frac{\sigma^2}{n} \frac{  2L(1/3L^2+1/2L+1/6)}{7/12L^2+1/2L+1/6} \nonumber
\end{align}

Simplifying the bias term in the MSE expansion: 
\begin{align}
    \sum^L_{\ell=1} (E[\hat{y_{\ell}}]-y_{\ell})^2
    &= \sum^L_{\ell=1} (E[\hat{y_{\ell}}]^2 -2E[\hat{y_{\ell}}]y_{\ell}+y_{\ell}^2) \nonumber
    \\
    &=  \sum^L_{\ell=1} E[\bar{\bar{y}}]^2-2E[\bar{\bar{y}}]E[\hat{\beta}]\Bar{x}+E[\hat{\beta}]^2\bar{x}^2+2E[\bar{\bar{y}}]E[\hat{\beta}]x_{\ell} \nonumber
    \\ 
    &
    -2E[\hat{\beta}]^2\Bar{x}x_{\ell}+E[\hat{\beta}]^2x_{\ell}^2-2E[\bar{\bar{y}}]E[y_{\ell}]+2E[\hat{\beta}]\Bar{x}E[y_{\ell}]-2E[\hat{\beta}]x_{\ell}E[y_{\ell}]+E[y_{\ell}^2] \nonumber
    \\
    &=  \sum^L_{\ell=1} E[\bar{\bar{y}}]^2-(L+1)E[\bar{\bar{y}}]E[\hat{\beta}]+\frac{(L+1)^2}{2}E[\hat{\beta}]^2+2E[\bar{\bar{y}}]E[\hat{\beta}]x_{\ell} \;\;\;\;\;\;\;\;\;\;\;\;\;\;\;\;\;\;\;\;\; \mbox{(substituting $\bar{x}=(L+1)/2$)} \nonumber
    \\ 
    &
    -(L+1)E[\hat{\beta}]^2x_{\ell}+E[\hat{\beta}]^2x_{\ell}^2-2E[\bar{\bar{y}}]E[y_{\ell}]+(L+1)E[\hat{\beta}]E[y_{\ell}]-2E[\hat{\beta}]x_{\ell}E[y_{\ell}]+E[y_{\ell}^2] \nonumber
    \\
    &= \frac{L\sigma^2}{n}+\frac{(2L^2+3L+1)}{6}  \left( \frac{ \sum^L_{\ell=1} (\ell-\frac{(L+1)}{2})(y_{\ell}-\bar{y})}{(1/12)(L^2-1)} \right)^2 \nonumber
\end{align}

\subsection{MSE in the model-free case}
Consider a problem with a sample size of $n$ and a number of levels $L$ where each level has an equal amount of samples (i.e., $n/L$), and all levels have the same variance (homoscedasticity).
Similarly, The mean squared error (MSE) for the model-free method is defined as:
\begin{equation}
    MSE = E[\sum^L_{\ell=1}(\hat{y_{\ell}}-y_{\ell})^2], \nonumber
\end{equation}
where $\hat{y_{\ell}}$ is the estimator for the true $y_{\ell}$ and $y_{\ell}$ is the true mean for $y_{\ell i}$ values (i.e., death estimates at level $\ell$).
\allowdisplaybreaks
\begin{align}
     MSE &= \sum^L_{\ell=1}( \underbrace{E [(\hat{y_{\ell}}-E[\hat{y_{\ell}}])^2]}_{\mbox{variance}}+\underbrace{E[(E[\hat{y_{\ell}}]-y_{\ell})^2]}_{\mbox{bias} = 0}) \;\;\;\;\;\;\;\;\;\;\;\; \mbox{model-free (direct) estimation $E[\hat{y_{\ell}}]=y_{\ell}$} \nonumber
     \\
     &= \sum^L_{\ell=1}(E[\hat{y_{\ell}}^2]-E[\hat{y_{\ell}}]^2) = \sum^L_{\ell=1}(E[L/n (\sum^{n/L}_{i=1} y_{\ell i})^2]-E[\hat{y_{\ell}}]^2) \nonumber
     \\
     &= \sum^L_{\ell=1}((L/n)^2(n/L (y_{\ell}^2+\sigma^2)+n/L(n/L-1) y_{\ell}^2)-y_{\ell}^2) = \frac{L^2\sigma^2}{n} \nonumber
\end{align}

\bibliographystyle{scsproc}

\bibliography{bib}

\section*{Author Biographies}
\vspace{-0.4em}

\textbf{\uppercase{Abdulrahman A. Ahmed}} is a PhD student in the Department of Industrial Engineering at University of Pittsburgh. He obtained his MSc and BSc in Operations Research and Computer Science from Cairo University. His current research is on developing methods that can get an accurate inference for complex sociotechnical systems with a focus on public health. His email address is \email{aba173@pitt.edu}.

\vspace{-0.25em}
\textbf{\uppercase{M. Amin Rahimian}} is an Assistant Professor in the Department of Industrial Engineering at University of Pittsburgh. His current research focus is on challenges of inference and intervention design in complex, large-scale sociotechnical systems with applications ranging from social networks and e-commerce to public health. His email address is \email{rahimian@pitt.edu}, and his website is \url{https://aminrahimian.github.io}.

\vspace{-0.25em}
\textbf{\uppercase{Mark S. Roberts}} is a distinguished professor in the Department of Health Policy and Management at University of Pittsburgh, and holds secondary appointments in Medicine, Industrial Engineering, Business Administration, and Clinical and Translational Science. His recent research has concentrated in the use of mathematical methods from operations research and management science, including Markov Decision Processes, Discrete Event, and Agent-based Simulation. His email address is \email{mroberts@pitt.edu} and his page is: \url{https://www.sph.pitt.edu/directory/mark-roberts}.

\end{document}